\documentclass{an}
\usepackage{graphicx}
\usepackage{times}
\usepackage{fancyhdr}
\def\mch{M$\rm^{c}$Hardy\,}
\newcommand{\uJy}{{\rm\thinspace \mu Jy}}
\newcommand{\etal}{{\it et al. }}
\begin{document}

\title{AGN in deep radio/X-ray surveys: hunting the earliest massive galaxies}

\author{Nick Seymour\inst{1}, Derek Moss\inst{2}, Ian \mch\inst{2}, 
  Katherine Gunn\inst{2}, Mat Page\inst{3}, Keith Mason\inst{3}, Nic 
  Loaring\inst{3}, Tom Dwelly\inst{3}}
\institute{Spitzer Science Center, California Institute of Technology, 
  Mail Code 220-6, 1200 East California Boulevard, Pasadena, CA 91125 USA.
  \and
  School of Physics \& Astronomy, University of Southampton, Highfield, 
  Southampton, SO17 1BJ, UK.
  \and
  MSSL, University College London, Dorking, Surrey, RH5 6NT, UK.
}

\date{Received; accepted; published online}

\abstract{Despite the plethora of deep (sub-mJy) radio surveys there remains 
  considerable doubt as to the exact nature of the galaxies contributing 
  to the source counts. Current evidence suggests that starformation in 
  moderately luminous {\it normal} galaxies is responsible for the bulk 
  of the emission below 1mJy. However given the sensitivities of these 
  surveys we would expect a fraction of these sources to be distant radio 
  galaxies. Using deep VLA and GMRT data we have found $\sim20$ high-z 
  candidate radio galaxies in two fields using the classical ultra-steep 
  radio spectrum technique (De Breuck \etal 2000) and selecting galaxies 
  with faint ($i'>25$) optical counterparts. Several of these sources have 
  X-ray detections in our deep {\it XMM/Chandra} observations and have 
  fluxes high enough to put them in the quasar regime if they lie above 
  redshift 3. Recently performed {\it Spitzer} GTO observations and upcoming
  near-infrared observations will help reveal the nature of these sources.
  \keywords{galaxies: active, formation, evolution}}

\correspondence{seymour@ipac.caltech.edu}

\maketitle

\section{Introduction}

Deep extra-galactic radio surveys have been detecting numerous galaxies at 
20cm with fluxes below 1mJy for 20 years now 
(Condon \etal 1989; Windhorst 
\etal 1990; Hopkins \etal 1998; Ciliegi \etal 1999; Gruppioni \etal 1999; 
Richards 2000; Prandoni \etal 2001; Bondi \etal 2003; Hopkins \etal 2003; 
Seymour \etal 2004). 
However understanding the nature 
of these sources has not progressed far. Classically radio surveys have 
detected quasars (and radio galaxies if the central black hole is hidden 
by dust in the optical) where the radio emission is often dominated by 
huge radio lobes now thought to be powered by jets. As surveys became more 
sensitive weak radio emission was found to be associated with starforming
galaxies (eg Condon 1992) and that it was correlated with the far-infrared 
emission (Carilli \& Yun, 2000). 

\subsection{Source counts}

Source counts are the most basic diagnostic of any survey of the sky and give 
us our first chance to interpret the data we have collected. The many surveys 
mentioned above all find an up-turn below 1mJy in the Euclidean normalised 
differential source counts above that expected from the extrapolation of the 
counts at higher flux densities. This has been interpreted by many authors as 
being due to a new population of {\it normal} star forming galaxies and 
there is some evidence that starformation is the dominant emission mechanism 
in the radio for several sub-mJy sources:

\begin{enumerate}

\item
Our two deep VLA 1.4GHz surveys (referred to as the 13hr and 1hr {\it 
  XMM/ROSAT} deep survey fields) have found $\sim450$ sources with flux 
densities greater than $30\uJy$ in each $30'$ diameter region. We have 
found the number of unambiguous Active Galactic Nuclei, AGN, (ie large 
radio lobes and/or broad optical emission lines) decreases regularly with 
decreasing flux density (Seymour \etal 2005, in prep.).

\item
Modeling of the source counts finds that the starforming population of 
galaxies needed to explain the sub-mJy upturn must evolve quickly with 
redshift (Seymour \etal 2005). This evolution is compatible with (and 
constrained 
by) the need for the starformation rate density to rise quickly from 
$z=0 - 1.0$ (Hopkins \etal 2004).

\item
Alfonso \etal 2005 have found direct evidence from spectral line ratios for 
starformation rates high enough to explain the observed radio luminosities 
of these faint galaxies. They also find that luminosity function of the 
starforming radio population evolves up to $z\sim0.5$ at rate consistent with 
the models and analysis (e.g. Haarsma \etal 2000; Hopkins 2004)

\end{enumerate}

\begin{figure}
\resizebox{\hsize}{!}
{\includegraphics[angle=270]{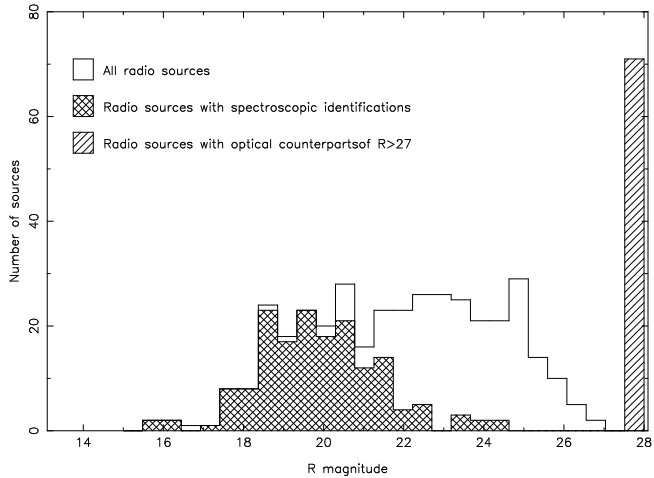}}
\caption{$R-$band distribution of the magnitude of optical counterparts to the 
  our 13hr {\it XMM/ROSAT} 1.4GHz survey, including the 165/449 that have 
  spectroscopically confirmed redshifts.}
\label{fig:op20}
\end{figure}

However the identifications of these radio sources are limited by their 
optical magnitudes. Figure~\ref{fig:op20} shows the distribution of $R-$band 
magnitudes of optical counterparts to sources from the 13hr field. As can 
been seen there is a continuous distribution sources from $R\sim18-25$ 
before a tail off due to optical incompleteness. Around $20\%$ of sources 
remain un-identified, similar to other comparatively deep surveys. Also 
shown in Fig.~\ref{fig:op20} is the fraction sources with spectroscopic 
identifications, less than half of the total sample. Other published 
surveys have similar spectroscopic detection rates. Therefore the 
current conclusions 
on the nature of the faint radio source population are drawn from only 
the optically brighter sources.

\begin{figure}
\resizebox{\hsize}{!}
{\includegraphics[angle=270]{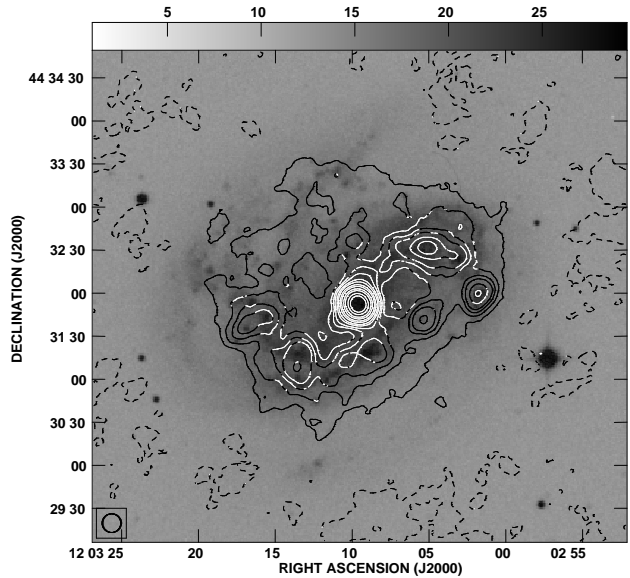}}
\caption{NGC 4051 at 20cm (contours) from a combination of VLA C 
  \& D Array overlaid on a Digital Sky Survey grayscale. It is clear 
  that as well as strong radio emission from the AGN there is significant
  contribution from starformation in the spiral arms.}
\label{fig:gal4051}
\end{figure}

An additional issue is that some sources may have contribution to the radio 
from both AGN and starformation as can be seen in the local universe.  
In Fig.~\ref{fig:gal4051} we can see radio contours overlayed onto an 
optical image of the Seyfert 1 galaxy NGC4051. As well as the core being 
very luminous in the radio, there is also more extended radio emission 
from the starformation in the spiral arms. Although in this case the AGN 
outshines the radio emission from starformation by an order of magnitude, 
it does not mean that in other galaxies with less active black holes and 
weaker radio jets that the AGN and starformation will not be radiating 
more equally.

\section{Our GMRT data}

Radio spectra provide a particularly useful
diagnostic.  Starburst emission is optically thin synchrotron
emission, having a normal ({\it i.e.} steep, $\alpha \sim -0.7$ where $S
\propto \nu^{\alpha}$) spectrum, and is also likely to be extended on
galaxy-size scales. Powerful radio AGN, with extended steep spectrum
radio lobes, are easily distinguished, but are rare in deep
surveys. More commonly, emission from faint AGN is point-like and of
relatively flat spectrum, perhaps from a compact jet. However some
older AGN will have diffuse steep spectrum lobes that will be brighter at
lower frequencies and hence be missed by high-resolution observations
at 20cm.

Therefore in august 2004 we obtained data on our 1hr and 13hr fields at 610MHz 
and 235MHz with the Giant Metrewave Radio Telescope (GMRT) in order to obtain
3-point radio spectral indices of our sources. From 4hours of observations, 
including calibration, 
we have obtained a map at 610MHz with a rms noise level of $57 \mu$Jy/beam 
at the phase centre with a dynamic range just below
1000 making it the deepest 610MHz extra-galactic survey to the best of 
our knowledge (Moss et al., in prep.).

\begin{figure}
\resizebox{\hsize}{!}
{\includegraphics[]{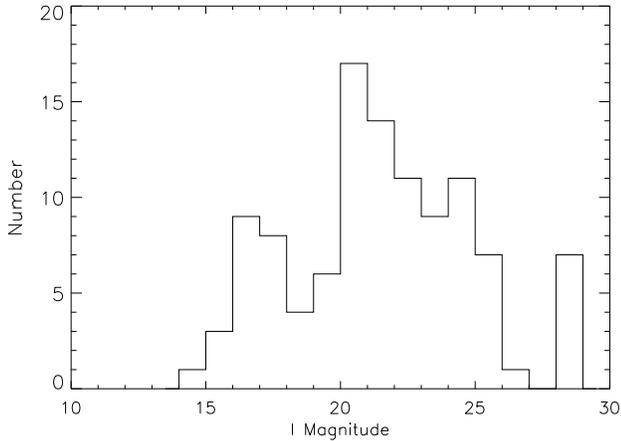}}
\caption{$i'-$band distribution of the magnitude of optical counterparts to 
the 1hr {\it XMM/ROSAT} 610MHz survey.}
\label{fig:op50}
\end{figure}

Over 200 sources are detected in each field at 610MHz within a diameter of 
$1^\circ$ with fluxes greater than $300\uJy$. Around 100 of these sources 
are also 
detected in the smaller 1.4GHz VLA maps. Most steep spectrum sources, i.e. 
$\alpha\sim-0.7$, are associated with optically bright spiral galaxies.
We have several ultra-steep spectrum sources, $\alpha\le -1$, with faint 
optical counterparts. The distribution of $i'-$band counterparts to 
these 610MHz sources is more bi-modal (see Fig.~\ref{fig:op50}) 
than the distribution of optical counterparts to the 1.4GHz surveys. 
This distribution is consistent with the observations that many sources are 
bright spirals, but what is the nature of the optically fainter sources?

\section{HzRG candidates}

Both radio galaxies and quasars (i.e. radio-loud QSOs) host massive
black holes and are very luminous at all wavelengths, but radio
galaxies are preferred for the study of galaxy formation at high redshifts
as quasars tend to out shine their host galaxies. In radio
galaxies the active nucleus is hidden by optically thick material just
beyond the accretion disk and so the host galaxy is detectable.
There is now considerable evidence that high-z radio galaxies (HzRGs)
are hosted by massive galaxies at every epoch (Archibald {\it et al.} 2001; 
Pentericci {\it et al.} 2001; Stern {\it et al.} 2003; Rocca-Volmerange {\it 
  et al.} 2004) 

High redshift radio galaxies are rare but, over the last decade or so,
it has become clear that ultra-steep radio spectra (USS; $\alpha <
-1.3$; $S \propto \nu^{\alpha}$) provide the best selection diagnostic of high
redshift radio galaxies (e.g. De Breuck {\it et al.} 2000). In the De Breuck
{\it et al.} sample $35\%$ of USS candidates are confirmed to have $z>3$. 
The De Breuck {\it et al.} sample has a flux limit of $S_{1.4GHz} > 10$mJy.
This sample preferentially samples high luminosities 
in the redshift range 1 to 5. But what will we find at lower flux
limits?

There are two main possibilities. We may find lower luminosity sources
at high redshifts, and so be able to study properly the evolution of
the radio luminosity function at high redshifts. We may also find
HzRGs at unprecedented redshifts (ie z $>6$) and so observe galaxies and
massive black holes in their earliest stages of formation.  Thus USS
samples selected at low (sub-mJy) flux limits have great potential for
the study of galaxy formation.

\subsection{Our Sample}

In each field we find around 10 sources with $\alpha \le -1.0$ and no 
confirmed optical counterparts, {\it i.e.} $i'>25$. These make up our 
candidate sample 
which include several objects also detected in our {\it XMM/Chandra} 
observations. If these sources are high-z radio galaxies then the X-ray 
emission is most likely from the accretion onto the central supermassive 
black hole. Whilst X-ray emission from the AGN is often absorbed in radio 
galaxies, mainly at harder energies $>1$keV, the absorption may be shifted 
out of the hard X-ray band at high redshifts, partly negating the decrease in 
flux density from the luminosity distance squared dimming.


\begin{figure}
\resizebox{\hsize}{!}
{\includegraphics[]{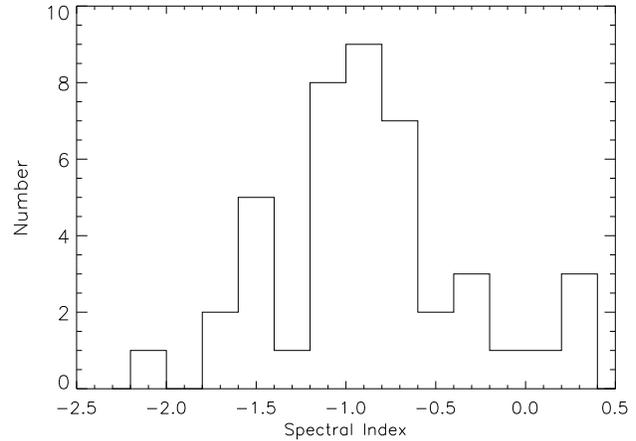}}
\caption{Distribution of radio spectral indices ($\alpha^{1.4GHz}_{610MHz}$, 
  $S_\nu \propto \nu^\alpha$) of sources common to the 1.4GHz VLA and 610MHz 
  GMRT surveys of the 1hr {\it XMM/ROSAT} field.}
\label{fig:alpha}
\end{figure}

\subsection{Expected Redshift - Luminosity Distribution of our sample}
\begin{figure}
\resizebox{\hsize}{!}
{\includegraphics[angle=270]{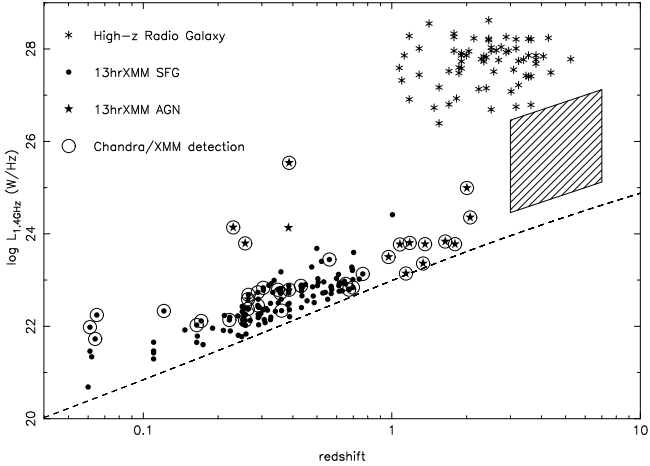}}
\caption{Distribution in luminosity/redshift space of the 165 radio sources 
  from our 13hr field with spectroscopic redshifts (stars = AGN, filled 
  circles = starforming galaxies, 
  circles = detection by {\it Chandra} and/or {\it XMM}). The sample of high-z 
  radio galaxies is that currently being studied by Spitzer with GO Cycle 1 
  data (PI Daniel Stern) and is indicative of the typical luminosity 
  distribution of HzRGs above $z=1$. The shaded region shows the expected 
  distribution of our candidate sources given the select effects (radio flux 
  and no optical ID) and the likely impact of the onset of reionisation at 
  $z=7$.}
\label{fig:lz}
\end{figure}

The redshift-luminosity (1.4GHz) distribution of 165/449 radio sources in 
our 13hr field, representing $\sim80\%$ of the sources with 
$R\le22$, is presented in Fig.~\ref{fig:lz} Confirmed AGN (broad emission 
lines and/or radio lobes) are plotted 
as stars and the rest, presumed to be star-forming galaxies although there 
are probable exceptions, are represented by filled circles. Sources circled 
have also been detected with our deep {\it XMM/Chandra} data (which is 
identical to that of the 1hr field). 
Plotted as reference is a sample of high redshift radio galaxies 
selected to cover the typical redshift-luminosity space of HzRGs and currently
being studied with Spitzer in a GO Cycle 1 proposal (PI Daniel Stern). 
The detection limit of $30\mu {\rm Jy}$ is indicated by the 
dashed line. The expected range of any detected high-z radio galaxies is 
indicated by the shaded region. The boundary in redshift is determined by 
$z=3$ cutoff due to USS selection and the faint optical nature of the sources, 
whilst the $z=7$ cutoff represents the approximate era of reionisation and the 
earliest a supermassive black-hole can have gained enough mass to be radiating 
at the proposed luminosities. The 
boundary in luminosity is determined simply as being 2 orders of magnitude 
from the detection limit at 1.4GHz (given the pre-selection of 
$S_{610MHz}\ge0.3$mJy), i.e. within the reasonable expectation of the value of 
the luminosities due to Malmquist bias. 

Figure~\ref{fig:lz} therefore shows that any high-z 
radio galaxies we detect will be either less luminous counterparts to the 
known HzRGs (although too luminous to be star-forming galaxies) or will be 
HzRGs at un-precedented redshifts likely accreting at 
high-rate as early super-massive black-holes are being formed. In the former 
case detecting less luminous counterparts to known HzRGs will increase the 
number of low-luminosity HzRGs improving the study of observables influenced 
by the AGN power (e.g. the amount of scattered light).
Additionally these observations would be vital in constraining the less 
luminous end of the radio AGN luminosity function (Dunlop \& Peacock 1990; 
Willot {\it et al.} 2001; Brookes \etal, these proceedings). In the 
latter case we may have detected spheroidal galaxies and super-massive black
holes co-forming (e.g. Magorrian {\it et al.} 1998; Tremaine {\it et al.} 
2002).

\section{Future work \& conclusions}

We have demonstrated the feasibility of using sub-mJy, multi-wavelength 
observations to select HzRG candidates using the VLA and GMRT. These 
sources may either be low luminosity
counterparts to the traditional family of HzRGs or HzRGs at unprecedented 
redshifts still in the early stages of formation. Either way these sources 
can provide vital clues to the early evolution of SMBH and giant elliptical 
galaxies. Clearly these sources 
warrant follow-up and through a combination of guaranteed UKIRT WFCAM 
and Palomar WIRC time we will obtain NIR imaging of these sources to 
$K\sim21$ in the coming semester. We have also applied for deeper GMRT 
time on each field to an 
rms sensitivity of $25\uJy$ at 610 MHz in order to use radio spectra to 
help us distinguish between AGN and starforming galaxies. These deeper 
observations will no doubt throw up more HzRG candidates. 

Additionally in June of this year the 13hr field was observed at Infrared 
wavelengths with the IRAC and MIPS instruments aboard {\it Spitzer} as 
part of the instrument teams' GTO time (PIs Rieke and Fazio). These 
observations will be of an equivalent depth to the 4th deepest {\it Spitzer}
survey. These data are clearly crucial in probing the restframe optical 
NIR emission of these potential HzRGs and determining initial photometric 
redshifts. Spectroscopic redshifts could later be obtained through either NIR 
spectroscopy if bright enough or through the IRS instrument on {\it Spitzer} 
({\it e.g.} Houck \etal 2005).

\acknowledgements
{We would like to thank Montse, Rosa, Enrique and Jose Luis for organising a 
  superb conference in Granada. We would also like to thank the staff of the 
  GMRT for helping us in obtaining the data. In particular we give thanks to 
  Niruj Mohan looking after us in India. As always the Belgian deserves 
  thanks for much useful advice.}


\end{document}